\documentstyle[12pt,aps]{revtex}
\newcommand{\bold}[1]{\mbox{\boldmath $#1$}}
\newcommand{\dd}{{\rm d}}
\begin{document}
\tighten
\draft

\title{\flushleft Response to Compagno and Persico\footnote
{This article 
is written by V Hnizdo in his private capacity. 
No official support or endorsement by 
the Centers for Disease Control and Prevention
is intended or should be inferred.}}

\author{\flushleft V Hnizdo}

\address{\flushleft National Institute for Occupational Safety
and Health, 1095 Willowdale Road, Morgantown,\\ WV 26505, USA}                                      
\maketitle
\vspace{3ex}
\noindent{\bf Abstract}\\
Some mathematically incorrect claims of Compagno and Persico
in their reply (2002 {\it J.\ Phys.\ A: Math.\ Gen.} {\bf 35} 8965)
to my comment on their recent paper on self-dressing and radiation 
reaction in classical electrodynamics are pointed out.

\vspace{4ex}

\noindent
Compagno and Persico (CP) have replied \cite{Reply} to my comment 
\cite{Comment} on their
paper on self-dressing and radiation reaction in classical
electrodynamics \cite{CP}. CP 
acknowledge the main point of the comment, namely that the expression
for the time-averaged electromagnetic self-force obtained in \cite{VH1} 
for the test charge of a Bohr--Rosenfeld field-measurement procedure
and rejected in \cite{CP2} as incorrect can be obtained also using a
formula for the self-force which they  derived by different means
in \cite{CP}. In view of this fact, CP now endorse the expression 
in question as correct.
However, some claims in their reply call for my response.

The expression for the time-averaged self-force on a spherical uniform
charge $q$ of radius $a$ obtained in \cite{VH1} reads 
\begin{equation}
\bar{F}=\frac{q^2}{T}\int_0^T\dd t'\,Q(t')f(t')
\label{Fb}
\end{equation}
where
\begin{equation}
f(t')=-\frac{1}{2a^3}(2-\chi)(2-2\chi-\chi^2)\Theta(2-\chi)
\;\;\;\;\;\;\;\;\;\chi=\frac{T-t'}{a}.
\label{f}
\end{equation} 
Here, the speed of light $c=1$ and $Q(t)$ is the charge's one-dimensional 
trajectory, which is subject to
the conditions that $Q(t)=0$ for $t\le 0$ and $|Q(t)|\ll a$,
$|\dd Q(t)/\dd t|\ll c$ for $0<t<T$; $\Theta(x)$ is the Heaviside step function.
Instead of the simple closed-form expression  (\ref{f}) 
for the function $f(t')$, CP counter-proposed in \cite{CP2} the expression
(normalized here to conform with (\ref{f})):
\begin{equation}
f(t')=-\frac{2}{3 V^2}\sum_{n=0}^{\infty}\frac{(-1)^n}{n!}
[\delta^{(n+1)}(T-t')-\delta^{(n+1)}(-t')]\langle r^{n-1}\rangle
\;\;\;\;\;\;\;\;V=\case{4}{3}\pi a^3
\label{fCP}
\end{equation}
where
\begin{equation}
\langle r^{n-1}\rangle
=\int_{|\bold{r}_1|<a}\dd\bold{r}_1\int_{|\bold{r}_2|<a}\dd \bold{r}_2
\,|\bold{r}_2-\bold{r}_1|^{n-1}=\frac{72 V^2(2a)^{n-1}}{(n+5)(n+3)(n+2)}.
\label{r}
\end{equation}
(The above closed-form expression for $\langle r^{n-1}\rangle$ was given 
subsequently in \cite{VH2}.)
My first point here has to be unfortunately of a rather trivial nature.
The sign of expression (\ref{fCP}) is correct, if, as in the standard
notation of elementary calculus, 
$u^{(n)}[v(x)]\equiv\dd^n u(y)/\dd y^n|_{y=v(x)}$.
Thus, e.g., $\delta^{(n)}(T-t)\equiv\dd^n\delta(x)/\dd x^n|_{x=T-t}$.
With this notation, there is no sign misprint in the function $f(t')$ in
\cite{CP2} that CP now want to correct---but if  
$\delta^{(n)}(T-t)$ meant instead $\dd^n\delta(T-t)/\dd t^n=
(-1)^n\dd^n\delta(x)/\dd x^n|_{x=T-t}$, as CP suggest, then  
not only the sign but also the factor $(-1)^n$ would be there in error. 
In either case, the function $f(t')$ is now given incorrectly by the
expression (3) of \cite{Reply}---it either has the wrong overall sign, 
or the factor $(-1)^n$ there must be omitted.

Contrary to an assertion of CP, I have never claimed that the expression 
(\ref{fCP}) for the function $f(t')$ is incorrect.
I have rather pointed out in \cite{VH2} that, in order to obtain
the time-averaged self-force (\ref{Fb}), this expression was 
used incorrectly by CP in the requisite integration. Since this is an 
integration
with finite limits involving high-order derivatives of the delta function,  
it cannot be performed as simplistically as CP have done, obtaining
\cite{CP2}
\begin{equation}
\bar{F}_{\rm CP}=-\frac{2\rho^2}{3T}\sum_{n=0}^{\infty}\frac{(-1)^n }{n!}\,
[Q^{(n+1)}(T)-Q^{(n+1)}(0)]\langle r^{n-1}\rangle\;\;\;\;\;\;\;\;\;
\rho=\frac{q}{V}.
\label{FbCP}
\end{equation}
This is the `exact' expression for the time-averaged self-force $\bar{F}$
that CP still claim in \cite{Reply} to be correct, despite the fact that
it is inconsistent with their newly adopted approval of the expression 
(\ref{f}) for the function $f(t')$.
This can be shown by evaluating the time-averaging integral (\ref{Fb})
for $\bar{F}$  using the expression (\ref{f}) for $f(t')$ and
the Taylor expansion of the trajectory $Q(t)$ about the point $t=T$: 
\begin{eqnarray}
\bar{F}&=&\frac{q^2}{T}\sum_{n=0}^{\infty}\frac{(-1)^n }{n!}\,Q^{(n)}(T)
\int_0^T\dd t'(T-t')^n f(t')\nonumber \\
&=&\frac{q^2}{T}\sum_{n=0}^{\infty}\frac{(-1)^n }{n!}\,Q^{(n)}(T)
\left\{\alpha_n-\left[a^{n-2}\left(\frac{2\kappa^{n+1}}{n+1}-
\frac{3\kappa^{n+2}}{n+2}+\frac{\case{1}{2}\kappa^{n+4}}{n+4}\right)
+\alpha_n\right]\Theta(2-\kappa)\right\}
\label{FT}
\end{eqnarray}
where
\begin{equation}
\alpha_n=\frac{48(2a)^{n-2}n}{(n+1)(n+2)(n+4)}\;\;\;\;\;\;\;\;\;\;\;\;\;\;\;
\kappa=\frac{T}{a}.
\label{alphakappa}
\end{equation}
This is the correct expression for the time-averaged self-force in terms of  
the derivatives $Q^{(n)}(T)\equiv\lim_{t\rightarrow T^-}\dd^nQ(t)/\dd t^n$.
(Note for completeness that the correct expression for $\bar{F}$ in terms 
of the derivatives $Q^{(n)}(0)\equiv\lim_{t\rightarrow 0^+}\dd^nQ(t)/\dd t^n$
is given by equation (25) of \cite{VH2}.)
As $\alpha_{n}=\case{2}{3}n\langle r^{n-2}\rangle/V^2$,
using (\ref{FT}) we can write $\bar{F}$ for $T\ge 2a$ as
\begin{equation}
\bar{F}=\frac{q^2}{T}\sum_{n=0}^{\infty}\frac{(-1)^n }{n!}\,Q^{(n)}(T)\alpha_n
=-\frac{2\rho^2}{3T}\sum_{n=0}^{\infty}\frac{(-1)^n }{n!}\,Q^{(n+1)}(T)
\langle r^{n-1}\rangle\;\;\;\;\;\;\;\;\;\;\;T\ge 2a
\end{equation}
which happens to be equal to the first of the two terms in the expression 
(\ref{FbCP}) of CP for $\bar{F}$. 
This demonstrates that expression (\ref{FbCP}) is incorrect;
for $T\ge 2a$, (\ref{FbCP}) can be corrected by dropping its second term,
but for $T<2a$, no such simple correction is possible.  

Far from being `convenient' for
an `exact' evaluation of the time-averaged self-force $\bar{F}$, formula 
(\ref{fCP}) 
is a purely formal expression that has no practical application in an 
integration  with finite limits. Its use by CP has led to 
the erroneous expression (\ref{FbCP}) for $\bar{F}$; when CP use it in 
\cite{Reply}
to prove the equivalence of expressions (\ref{f}) and (\ref{fCP}), they revert
the Taylor expansions  $\sum_n(-1)^n\delta^{(n+1)}(t)r^n/n!$ 
to $\delta'(t-r)$ before performing the finite-limit integration.

So far, CP have responded to my criticism \cite{VH1} of their re-analysis
\cite{CP1} of the Bohr--Rosenfeld field-measurement procedure only by
making mathematically incorrect claims.
All these involved rather simple mathematical points about which there 
should have 
been no need of explicating since my paper \cite{VH2} of 2000 (where I derived
expression (\ref{f}) using elementary calculus of the delta function), if not
already since my comment \cite{VH1} of 1999 (where I used Fourier transform methods).
It is regrettable that such points have deflected from the interesting
issues of physics relating to the famous Bohr--Rosenfeld analysis.

\end{document}